\documentstyle[12pt]{article}
  
 \renewcommand{\title}[1]{\null\vspace{25mm}
   \noindent{\Large{\bf #1}}\vspace{10mm}
    }
 \newcommand{\authors}[1]{\noindent{\large #1}\vspace{20mm}
    }
 \newcommand{\address}[1]{{\center{\noindent #1\vspace{10mm}}
    }}
 \renewcommand{\abstract}[1]{\vspace{17mm}
  \noindent{\small{\em Abstract.} #1}\vspace{2mm}
   }     
 
 \newcommand{\en}{\begin{equation}} \newcommand{\dv}{\end{equation}} \newcommand{\de}{\partial}        
 
 \newcommand{\dms}{\partial_{\mu}}  
  
 \newcommand{\tr}{\begin{description}}
 \newcommand{\st}{\end{description}} 
 \newcommand{\pr}{\hspace{1mm}}     \newcommand{\ptr}{\hspace{3mm}}
 \newcommand{\pe}{\begin{eqnarray}} \newcommand{\se}{\end{eqnarray}}
  
 \newcommand{\el}{{\cal L}}
   \newcommand{\be}{\beta}   
 \newcommand{\ro}{\varrho}  
 \newcommand{\sed}{\begin{array}}  \newcommand{\os}{\end{array}}  
 \newcommand{\ee}{\varepsilon}
 \newcommand{\la}{\lambda} \newcommand{\si}{\sigma} \newcommand{\al}{\alpha}
 \newcommand{\da}{\delta}  
 \newcommand{\nm}{\nonumber}
  
 \newcommand{\pp}{\hspace{5mm}}\newcommand{\am}{A_\mu}

 \newcommand{\ga}{\gamma}

  \newcommand{\Tr}{\hbox{Tr}} 
 \newcommand{\bd}{\begin{Def}\hspace{-2mm}: \rm}

\begin{document}   \setcounter{table}{0}
 
 \begin{titlepage}
 \begin{center}
 \hspace*{\fill}{{\normalsize \begin{tabular}{l}
                              {\sf REF. TUW 99-24}\\
                              \end{tabular}   }}

 \title{A comment on the 4D antisymmetric tensor field model}

 \authors{J. Rant$^1$, M. Schweda and H. Zerrouki$^2$}    \vspace{-20mm}
       
 \address{Institut f\"ur Theoretische Physik, 
                          Technische Universit\"at Wien\\
      Wiedner Hauptstra\ss e 8-10, A-1040 Wien, Austria}
      
 \footnotetext[1]{Work supported in part by the "Fonds zur F\"orderung der
 Wissenschaflicher Forschung", under Project Grant Number P11354-PHY.}
  \footnotetext[2]{Work supported in part by the "Fonds zur F\"orderung der
 Wissenschaflicher Forschung", under Project Grant Number P13125-TPH.}
 \end{center} 
 \thispagestyle{empty}
 \abstract{We show the existence of a renormalizable local supersymmetry for 
           the gauge fixed action of the four dimensional antisymmetric tensor
           field model in a curved background quantized in a generalized axial gauge. 
           By using the technique of the algebraic renormalization procedure, 
           we prove the ultraviolet finiteness of the model to all orders of perturbation theory.}
 \end{titlepage}
 
 \section{Introduction}
 

In \cite{Joe} the authors have shown that the four dimensional antisymmetric 
tensor 
field model, quantized in a curved background admitting Killing 
vectors\footnote{ In fact the manifold was chosen to be asymptotically flat,
having trivial topology and admitting Killing vectors.}, was anomaly free and
finite to all orders of perturbation theory. 
In this work we generalize the results of \cite{Joe} to be valid for manifolds
not necessarily admitting Killing vectors.\\
In order to avoid the difficulties the authors met in \cite{Joe,Hassan}, we introduce
a vector field $n^\mu (x)$ which will play the role of a generalized axial gauge
vector in curved space-time. In fact, from the beginning we choose the 
manifold $\cal M$, on which the four dimensional antisymmetric tensor field model is
discussed, to have a trivial topology. In particular, this means that the gauge
vector field $n^\mu (x)$ can be chosen to be nowhere vanishing. \\[2mm]
In the present paper we show, using the algebraic renormalization techniques 
\cite{Piguet,Buch,Becchi},
that the model is anomaly free and finite to all orders of perturbation theory.
In section 2 we describe the model as well as its gauge fixing. In section 3 we 
display the superdiffeomorphisms transformations. Section 4 is devoted to the 
off-shell analysis of the theory and finally the stability as well as the anomaly analysis are
performed in section 5.

 \section{The model}
 
 We begin with the classical action of the four dimensional antisymmetric field model in
curved space-time: 
 \en S_{inv}={1\over 4}\int d^4x\,\ee^{\mu\nu\ro\si}F_{\mu\nu}^aB_{\ro\si}^a\ptr, 
     \label{Sinv}   \dv
 where $\cal M$ is a curved manifold endowed with the Euclidean metric 
$g_{\mu\nu}$.
 $B_{\ro\si}^a$ stands for the antisymmetric tensor field whereas
 $F_{\mu\nu}^a$ is the field strength given by
 \en F_{\mu\nu}^a=\dms A_\nu^a-\de_\nu\am^a+f^{abc}\am^bA_\nu^c \ptr, \dv
 where $\am^a$ is the gauge field. All fields are Lie algebra valued and belong to the
 adjoint representation of some compact semi-simple gauge group $G$ whose structure
 constants $f^{abc}$ are completely antisymmetric in their indices. The generators of 
 the Lie algebra are chosen to be anti-hermitian and fulfilling $[T^a,T^b]=f^{abc}T^c$
 and $\Tr(T^aT^b)=\da^{ab}$.  Finally,
 $\ee^{\mu\nu\ro\si}$ is the totally antisymmetric Levi-Civita 
 tensor density\footnote{
 We denote by $g^{\mu\nu}$ the inverse of the metric and its determinant by $g$. Under
 diffeomorphisms, $\sqrt{g}$ behaves like a scalar density of weight +1 and the volume
 element $d^4x$ has weight $-1$. The Levi-Civita tensor density $\ee^{\mu\nu\ro\si}$ has
 weight +1 and its inverse 
 $$ \ee_{\mu\nu\ro\si}={1\over g}g_{\mu\al}g_{\nu\be}g_{\ro\ga}g_{\si\da}\ee^{\al\be\ga\da}$$
 carries weight $-1$. 
 } of weight +1.
 \\[2mm]
 The action (\ref{Sinv}) is invariant under the following two infinitesimal 
 symmetries:
 \pe \da^{(1)}\am^a &=& -\dms\theta^a-f^{abc}\am^b\theta^c\equiv -(D_\mu\theta)^a \ptr, \nm\\
     \da^{(1)}B_{\mu\nu}^a&=& f^{abc}\theta^bB_{\mu\nu}^c \ptr, \label{1sym}
     \se    
 and
 \pe \da^{(2)}\am^a &=& 0 \ptr, \nm \\
     \da^{(2)}B_{\mu\nu}^a &=& -(D_\mu\varphi_\nu-D_\nu\varphi_\mu)^a \ptr, \label{2sym}\se  
 where $\theta^a$ is the local gauge parameter
 and $\varphi^a_\mu$ is a local vector parameter. $D_\mu$ represents the 
 covariant derivative. 
In order to fix the gauge consistently we use a generalized axial\footnote{For different applications
of the non-covariant gauges in flat space-time in the context of the algebraic renomalization see
\cite{Buch}. In the present work, however, we generalize the axial gauge to curved manifolds having
a trivial topology.} gauge type 
with a local vector $n^\mu (x)$. In fact, we will quantize the model on a four dimensional
manifold which is assumed to be topologically trivial and asymptotically flat. Therefore, 
we can choose $n^\mu (x)$ to be a nowhere vanishing local vector.
Hence, the gauge fixing part of the action, which is metric dependent and therefore destroys the 
topological character of the model, reads:
 \en S_{gf}=s\int d^4x\sqrt g\left(\bar c^ag^{\mu\nu}n_\mu A_\nu^a+g^{\mu\al}g^{\nu\be}
     \bar\xi^a_\be n_\al B_{\mu\nu}^a+\bar\phi^ag^{\mu\nu}n_\mu\xi^a_\nu\right) \ptr,
     \label{Sgf} \dv
 where the vector $\xi_\mu^a$ is the ghost field for the symmetry (\ref{2sym}), $\phi^a$ is
 the ghost for the ghost $\xi_\mu^a$ and $c^a$ is the ghost for the symmetry (\ref{1sym}).
 We collect the antighosts and the corresponding Lagrange multipliers in pairs 
 $(\bar c^a,b^a),(\bar\xi_\mu^a,h_\mu^a)$ and $(\bar\phi^a,\omega^a)$. \\
Contrary to \cite{Joe}, gauge-fixing the four dimensional antisymmetric tensor field 
model using the generalized axial gauge is much simpler than using the 
Landau gauge such that (\ref{Sgf}) takes a simple form (see the expression (2.19) in \cite{Joe}). 
In the present case the extended nilpotent BRS-transformations read as
 \pe s\am^a &=& -(D_\mu c)^a \ptr, \nm \\
     sB_{\mu\nu}^a &=& -(D_\mu\xi_\nu-D_\nu\xi_\mu)^a+f^{abc}c^bB_{\mu\nu}^c-
     \ee_{\mu\nu\ro\si}f^{abc}\sqrt gg^{\ro\al}g^{\si\be}n_\al\bar\xi_\be^b\phi^c\ptr, \nm \\
     s\xi_\mu^a &=& (D_\mu\phi)^a+f^{abc}c^b\xi_\mu^c\ptr, \nm \\
     s\phi^a &=& f^{abc}c^b\phi^c \nm \ptr, \\
     sc^a &=& {1\over 2}f^{abc}c^bc^c \nm \ptr, \\
     s\bar c^a &=& b^a \pr, \hspace{1cm} sb^a=0 \ptr, \nm \\
     s\bar\xi_\mu^a &=& h_\mu^a\pr, \hspace{1cm} sh_\mu^a=0 \ptr, \nm \\
     s\bar\phi^a &=& \omega^a \pr, \hspace{1cm} s\omega^a=0 \ptr, \nm \\
     sg_{\mu\nu}&=&\hat g_{\mu\nu} \pr, \hspace{1cm} s\hat g_{\mu\nu}=0 \ptr. 
     \label{BRS} \se 
 The metric plays in (\ref{Sgf}) the role of a gauge parameter \cite{Joe} which we also let transform
 as a BRS-doublet as given in the last line of (\ref{BRS}). 
 Furthermore, to control the $n_\mu$-dependence of the theory we use the arguments of \cite{ndep2}
and enlarge the 
 BRS-transformations  by allowing also a variation of the local vector
 $n_\mu$:
 \en s¸n_\mu=\chi_\mu\pr,\pp s\chi_\mu=0\ptr, \label{BRSn} \dv
 and add the following term to the action
 \en S_n=-\int d^4x\sqrt g\left(\bar c^ag^{\mu\nu}\chi_\nu\am^a+g^{\mu\al}g^{\nu\be}
     \bar\xi_\be^a\chi_\al B_{\mu\nu}^a-\bar\phi^ag^{\mu\nu}\chi_\nu\xi_\mu^a
     \right) \ptr. \dv
 Here, $\chi_\mu$ is a local anticommuting vector parameter. It turns out that the BRS-operator is
 nilpotent on-shell\footnote{It should be mentioned that contrary to \cite{Joe} our analysis using
the generalized axial gauge gets simpler due to the fact that we have less fields.}:
 \en s^2B_{\mu\nu}^a=-\ee_{\mu\nu\ro\si}f^{abc}{\da(S_{inv}+S_{gf}+S_n)\over\da B_{\ro\si}^b}
     \phi^c \pp \hbox{and}\pp s^2=0 \pp \hbox{for all other fields.} \dv
 One can easily verify the BRS-invariance of the gauge fixed action, which obeys
 \en s(S_{inv}+S_{gf}+S_{n})=0 \ptr. \dv
 We present the canonical dimensions and the Faddeev--Popov charges of all fields in 
 Table 1. 
 \begin{center}
 \begin{table}[h] \label{fields}
 \begin{center} 
 \begin{tabular}{|c|c|c|c|c|c|c|c|c|c|c|c|c|c|c|c|} 
 \hline
  & $\am^a$ & $B_{\mu\nu}^a$ & $\phi^a$ & $\xi_\mu^a$ & $c^a$ & $\bar\xi_\mu^a$ & $\bar c^a$ 
  & $\bar\phi^a$ & $b^a$ & $h_\mu^a$ & $\omega^a$ & $n_\mu$ & $\chi_\mu$ & $g_{\mu\nu}$ 
  & $\hat g_{\mu\nu}$ \\
 \hline
 dim & 1 & 2 & 0 & 1 & 0 & 2 & 3 & 3 & 3 & 2 & 3 & 0 & 0 & 0 & 0 \\
 \hline
 $\phi\pi$ & 0 & 0 & 2 & 1 & 1 & -1 & -1 & -2 & 0 & 0 & -1 & 0 & 1 & 0 & 1\\
 \hline 
 \end{tabular} 
 \caption{Dimensions and Faddeev--Popov charges of the fields}
 \end{center}
 \end{table}
 \end{center}
 
\section{Superdiffeomorphisms} 
 
 As already shown in \cite{Joe} for the Landau-type gauge, the four dimensional
 antisymmetric tensor field model possesses 
besides the BRS-symmetry and the invariance under diffeomorphisms a further 
invariance
 of supersymmetric-kind, namely the so-called superdiffeomorphisms. For these local
 transformations we propose:
  
 \pe \da_{(\eta)}\am^a &=& \ee_{\mu\nu\ro\si}\eta^\nu\sqrt gg^{\ro\al}g^{\si\be}n_\al
     \bar\xi^a_\be \ptr, \nm \\
     \da_{(\eta)}B_{\mu\nu}^a &=& \ee_{\mu\nu\ro\si}\eta^\ro\sqrt gg^{\si\al}n_\al\bar c^a
     \ptr, \nm \\
     \da_{(\eta)}c^a &=& -\eta^\mu\am^a \ptr, \nm \\
     \da_{(\eta)}\bar c^a &=& 0 \ptr, \nm \\
     \da_{(\eta)}b^a &=& \el_\eta\bar c^a \ptr, \nm \\
     \da_{(\eta)}\xi^a_\mu &=& \eta^\nu B_{\mu\nu}^a \ptr, \nm \\
     \da_{(\eta)}\bar\xi^a_\mu &=& -g_{\mu\nu}\eta^\nu\bar\phi^a \ptr, \nm \\
     \da_{(\eta)}h_\mu^a &=& \el_\eta\bar\xi^a_\mu+s(g_{\mu\nu}\eta^\nu\bar\phi^a)\ptr, \nm \\
     \da_{(\eta)}\phi^a &=& \eta^\mu\xi_\mu^a \ptr, \nm \\
     \da_{(\eta)}\bar\phi^a &=& 0 \ptr, \nm \\
     \da_{(\eta)}\omega &=& \el_\eta\bar\phi^a \ptr, \nm \\
     \da_{(\eta)}n_\mu &=& 0\ptr, \nm \\
     \da_{(\eta)}\chi_\mu &=& \el_\eta n_\mu\ptr, \nm \\
     \da_{(\eta)}g_{\mu\nu} &=& 0 \ptr, \nm \\
     \da_{(\eta)}\hat g_{\mu\nu} &=& \el_\eta g_{\mu\nu} \ptr, \label{sdiff} \se
 where $\el_\eta$ represents the Lie derivative and $\eta^\mu$ is the vector parameter of
 the transformations carrying ghost number +2. The resulting algebra between the BRS-operator
 and the superdiffeomorphisms closes on-shell:
 \pe \{s,\da_{(\eta)}\} &=& \el_\eta+ \hbox{ equations of motion}  \ptr, \nm \\
     \{\da_{(\eta)},\da_{(\eta')}\} &=& 0 \ptr. \label{algebra}\se
 At this stage one remarks that contrary to the case of \cite{Joe} there is no constraint which
 requires the manifold to possess Killing vectors. Therefore, the underlying paper is a 
generalization of \cite{Joe}. 
 
\section{The off--shell analysis} 
  
 In order to describe the BRS-symmetry content consistently at the functional level,
 we introduce a set of external sources\footnote{One has to note that the sources have weight +1.} 
coupled
 to the nonlinear BRS-variations of the quantum fields:
 \pe S_{ext} &=& \int d^4x\,\left[\ga^{\mu\nu a}(sB_{\mu\nu}^a)+\Omega^{\mu a}(s\am^a)+
     L^a(sc^a)+D^a(s\phi^a)+\ro^{\mu a}(s\xi_\mu^a)\right]+\nm \\
     &+& {1\over 2}\int d^4x\,\ee_{\mu\nu\ro\si}f^{abc}\ga^{\mu\nu a}\ga^{\ro\si b}
     \phi^c \ptr. \se
 We display the canonical dimensions and the Faddeev--Popov charges of the external sources
 in Table 2.
 \begin{center}
 \begin{table}[ht] \label{sources} \begin{center} 
 \begin{tabular}{|c|c|c|c|c|c|} 
 \hline
  & $\ga^{\mu\nu a}$ & $\Omega^{\mu a}$ & $L^a$ & $D^a$ & $\ro^{\mu a}$  \\
 \hline
 dim & 2 & 3 & 4 & 4 & 3  \\
 \hline
 $\phi\pi$ & -1 & -1 & -2 & -3 & -2  \\
 \hline 
 \end{tabular} 
 \caption{Dimensions and Faddeev--Popov charges of the external sources}
 \end{center}
 \end{table}
 \end{center}
 Therefore, the complete action
 \en \Sigma=S_{inv}+S_{gf}+S_n+S_{ext} \label{Sigma} \dv
 obeys the Slavnov identity:
 \pe {\cal S}(\Sigma) &=&
     \int d^4x\,\left[{\da\Sigma\over\da\ga^{\mu\nu a}}{\da\Sigma\over\da B_{\mu\nu}^a}
     +{\da\Sigma\over\da\Omega^{\mu a}}{\da\Sigma\over\da\am^a}+
     {\da\Sigma\over\da L^a}{\da\Sigma\over\da c^a}+{\da\Sigma\over\da D^a}{\da
     \Sigma\over\da\phi^a}+{\da\Sigma\over\da\ro^{\mu a}}{\da\Sigma\over\da\xi_\mu^a}+
     \nm \right.\\  &+& \left. h_\mu^a{\da\Sigma\over\da\bar\xi_\mu^a}+ 
     b^a{\da\Sigma\over\da\bar c^a}+\omega^a{\da\Sigma\over\da\bar\phi^a}+
     \hat g_{\mu\nu}{\da\Sigma\over\da g_{\mu\nu}}+\chi_\mu{\da\Sigma\over\da n_\mu}\right]
      =0 
     \ptr.   \label{Slavnov} \se 
 For later use we introduce the linearized Slavnov operator ${\cal S}_\Sigma$:
  \pe {\cal S}_\Sigma &=&
     \int d^4x\,\left[{\da\Sigma\over\da\ga^{\mu\nu a}}{\da\over\da B_{\mu\nu}^a}
     +{\da\Sigma\over\da B_{\mu\nu}^a}{\da\over\da\ga^{\mu\nu a}}
     +{\da\Sigma\over\da\Omega^{\mu a}}{\da\over\da\am^a}
     +{\da\Sigma\over\da A_{\mu}^a}{\da\over\da\Omega^{\mu a}}
     + \right. \nm \\
     &+& \left.{\da\Sigma\over\da L^a}{\da\over\da c^a}+{\da\Sigma\over\da c^a}{\da\over\da L^a}
     +{\da\Sigma\over\da D^a}{\da\over\da\phi^a}+{\da\Sigma\over\da\phi^a}{\da\over\da D^a}
     +{\da\Sigma\over\da\ro^{\mu a}}{\da\over\da\xi_\mu^a}
     +{\da\Sigma\over\da\xi_\mu^a}{\da\over\da\ro^{\mu a}}+
     \nm \right.\\  &+& \left. h_\mu^a{\da\over\da\bar\xi_\mu^a}+ 
     b^a{\da\over\da\bar c^a}+\omega^a{\da\over\da\bar\phi^a}+
     \hat g_{\mu\nu}{\da\over\da g_{\mu\nu}}+\chi_\mu{\da\over\da n_\mu}\right]
       \ptr. \label{lin}
         \se 
The introduction of external sources leads to a linearly broken 
Ward identity for the superdiffeomorphisms: 
 \en {\cal W}^S_{(\eta)}\Sigma=\Delta_{(\eta)}^{cl} \ptr, \label{Wmu}\dv
 where
 \pe {\cal W}_{(\eta)}^S &=& \int d^4x\left[\ee_{\mu\nu\ro\si}\eta^\nu(\sqrt gg^{\ro\al}
     g^{\si\be}n_\al\bar\xi_\be^a-\ga^{\ro\si a}){\da\over\da A_\mu^a}
     -\eta^\mu\am^a{\da\over\da c^a}+\el_\eta\bar c^a{\da\over\da b^a}+  \right. \nm \\
     &+& \ee_{\mu\nu\ro\si}\eta^\ro(\sqrt gg^{\si\al} n_\al\bar c^a-\Omega^{\si a})
     {\da\over\da B_{\mu\nu}^a}+\eta^\nu B_{\mu\nu}^a{\da\over\da\xi_\mu^a}+
     \left(\el_\eta\bar\xi^a_\mu+s(g_{\mu\nu}\eta^\nu\bar\phi^a)\right)
     {\da\over\da h_\mu^a}+\nm\\ &+& \eta^\mu\xi_\mu^a{\da\over\da\phi^a}- 
      g_{\mu\nu}\eta^\nu\bar\phi^a{\da\over\da\bar\xi_\mu^a}+
     \el_\eta\bar\phi^a{\da\over\da\omega^a}+\el_\eta g_{\mu\nu}{\da\over\da 
     \hat g_{\mu\nu}}-\eta^\mu L^a{\da\over\da\Omega^{\mu a}}-
     \eta^\mu D^a{\da\over\da\ro^{\mu a}}- \nm \\ &-&   
      \left. \eta^\mu\ro^{\nu a}{\da\over\da\ga^{\mu\nu a}}+\el_\eta n_\mu{\da\over\da\chi_\mu}
      \right]\ptr \se
is the Ward operator for superdiffeomorphisms and
\pe \Delta_{(\eta)}^{cl} &=& \int d^4x\left[-\ga^{\mu\nu a}\el_\eta B_{\mu\nu}^a  
     -\Omega^{\mu a}\el_\eta A_\mu^a+L^a\el_\eta c^a-D^a\el_\eta\phi^a+
     \ro^{\mu a}\el_\eta\xi_\mu^a+ \nm \right. \\
     &+& \left. \ee_{\mu\nu\ro\si}\Omega^{\mu a}\eta^\nu s(\sqrt gg^{\ro\al}g^{\si\be}n_\al
     \bar\xi^a_\be)+
     \ee_{\mu\nu\ro\si}\ga^{\mu\nu a}\eta^\ro s(\sqrt gg^{\si\al}n_\al\bar c^a)     
     \right] \ptr \label{breaking}\se
is the breaking which is linear in the quantum fields and therefore
 harmless at the quantum level. \\
 On the other hand, if the functional $\Sigma$ is a solution of the Slavnov identity (\ref{Slavnov}),
 of the superdiffeomorphisms Ward identity (\ref{Wmu}) as well as  
 the Ward identity for diffeomorphisms
 \en {\cal W}^D_{(\ee)}\Sigma=0 \ptr, \label{P}\dv
 where ${\cal W}^D_{(\ee)}$ stands for the corresponding Ward operator
 \en {\cal W}^D_{(\ee)}=\int d^4x\sum_\varphi(\el_\ee\varphi){\da\over\da\varphi}\ptr,\dv
 for all fields $\varphi$, then the following off--shell algebra holds:
 \pe \{{\cal S}_{\Sigma},{\cal S}_{\Sigma}\} &=& 0 \nm \ptr, \\
     \left\{{\cal W}^S_{(\eta)},{\cal W}^S_{(\eta')}\right\} &=& 0 \nm \ptr, \\
     \left\{{\cal W}^D_{(\ee)},{\cal W}^D_{(\ee')}\right\} &=& -{\cal W}^D_{(\{\ee,\ee'\})}
     \ptr, \nm
  \\ \left\{{\cal S}_{\Sigma},{\cal W}^S_{(\eta)}\right\} &=& {\cal W}^D_{(\eta)} \nm \ptr, \\
     \left\{{\cal W}^D_{(\ee)},{\cal W}^S_{(\eta)}\right\} &=& -{\cal W}^S_{([\ee,\eta])}
     \nm \ptr, \\
     \left\{{\cal S}_{\Sigma},{\cal W}^D_{(\ee)}\right\} &=& 0 \ptr. \se
 Here, we used the  graded Lie brackets:
 \pe \{\ee,\ee'\}^\mu &=& \el_\ee\ee'^{\mu} \ptr,\nm \\
     \left[\ee,\eta\right]^\mu &=& \el_\ee\eta^\mu \ptr. \se
 It is straightforward to convince oneself that the total action (\ref{Sigma}) fulfills
 the gauge conditions
 \pe {\da\Sigma\over\da b^a} &=& \sqrt gg^{\mu\al}n_\al\am^a \ptr,\nm \\
     {\da\Sigma\over\da h_\mu^a} &=& -\sqrt gg^{\mu\al}g^{\nu\be}n_\nu B_{\al\be}^a \nm\ptr,\\
     {\da\Sigma\over\da\omega^a} &=& \sqrt gg^{\mu\al}n_\al\xi_\mu^a \ptr, \label{Eichbed}\se
 the following antighost equations
 \pe {\da\Sigma\over\da\bar c^a}+\sqrt gg^{\mu\al}n_\al{\da\Sigma\over\da\Omega^{\mu a}} &=&
     -s(\sqrt gg^{\mu\al}n_\al)\am^a \ptr, \nm \\
     {\da\Sigma\over\da\bar\xi_\mu^a}-\sqrt gg^{\mu\al}g^{\nu\be}n_\nu
     {\da\Sigma\over\da \ga_{\al\be}^a} &=& s(\sqrt gg^{\mu\al}g^{\nu\be}n_\nu)B_{\al\be}^a
     \ptr, \nm \\
     {\da\Sigma\over\da \bar\phi^a}-\sqrt gg^{\mu\al}n_\al{\da\Sigma\over\da\ro^{\mu a}}&=&
     s(\sqrt gg^{\mu\al}n_\al)\xi_\mu^a \ptr, \label{anti}\se
 and a further integrated constraint, namely the ghost equation
 \en {\cal G}^a\Sigma=\Delta^a \ptr, \label{ghosteq}\dv
 where
 \en {\cal G}^a=\int d^4x\left({\da\over\da\phi^a}-f^{abc}\bar\phi^b{\da\over\da b^c}
     \right) \ptr, \dv
 and 
 \en \Delta^a=\int d^4xf^{abc}\left(\sqrt g\ee_{\mu\nu\ro\si}g^{\ro\al}g^{\si\be}
     n_\al\ga^{\mu\nu b}\bar\xi_\be^c+D^bc^c+\ro^{\mu b}\am^c+{1\over 2}\ee_{\mu\nu\ro\si}
     \ga^{\mu\nu b}\ga^{\ro\si c}\right) \ptr. \dv
Here, $\Delta^a$ is a linear breaking.

\section{Proof of the finiteness}

 This section is devoted to discuss the full symmetry content of the theory at the 
 quantum level, i.e. the question of possible anomalies and the stability problem
 which amounts to analyze all invariant counterterms.  
 \\[2mm]
 We begin by studying the stability where 
 in the first step we consider one-loop corrections.
 This requires the analysis of the most general counterterms for the total
 action and implies to consider the following perturbed action
 \en \Sigma'=\Sigma+ \hbar \Delta \ptr, \dv
 where $\Sigma$ is the total action (\ref{Sigma}) and $\Sigma'$ is an arbitrary
 functional depending via $\Delta$ 
 on the same fields as $\Sigma$ and satisfying the Slavnov identity 
 (\ref{Slavnov}), the Ward identity for the
 superdiffeomorphisms (\ref{Wmu}), the gauge conditions (\ref{Eichbed}), the 
 antighost equations (\ref{anti}), the ghost equation (\ref{ghosteq}) and the Ward
 identity for the diffeomorphisms (\ref{P}). The 
 perturbation $\Delta$ collecting all appropriate invariant counterterms is an integrated
 local field polynomial of dimension four and ghost number zero. 
 \\[2mm]
 Now we are searching for the most general deformation of the classical action such
 that the perturbed action $\Sigma'$ still fulfills the above constraints. Therefore the 
 perturbation $\Delta$ has to obey the following set of equations:
 \pe {\da\Delta\over\da b^a} &=& 0 \ptr,  \label{gc1} \\
     {\da\Delta\over\da h_\mu^a} &=& 0 \ptr,  \label{gc2} \\
     {\da\Delta\over\da\omega^a} &=& 0 \ptr,  \label{gc3} \\
     {\da\Delta\over\da\bar c^a}+\sqrt gg^{\mu\al}n_\al{\da\Delta\over\da\Omega^{\mu a}} &=& 0
     \ptr,\label{a1}\\   
     {\da\Delta\over\da\bar\xi_\mu^a}-\sqrt gg^{\mu\al}g^{\nu\be}n_\nu
     {\da\Delta\over\da \ga_{\al\be}^a} &=& 0\ptr, \label{a2}\\
     {\da\Delta\over\da \bar\phi^a}-\sqrt gg^{\mu\al}n_\al{\da\Delta\over\da\ro^{\mu a}} &=& 0
     \ptr, \label{a3} \\
     {\cal S}_{\Sigma}\Delta &=& 0 \ptr,  \label{con1} \\
     {\cal W}_{(\eta)}^S\Delta &=& 0 \ptr,  \\
     {\cal W}_{(\ee)}^D\Delta &=& 0 \ptr,\label{con2} \\
     \int d^4x {\da\Delta\over\da\phi^a} &=& 0 \ptr. \label{ghost234}    
      \se
 The first three equations (\ref{gc1})--(\ref{gc3}) imply that the  
 perturbation $\Delta$
 does not depend on the multiplier fields $b^a$, $h_\mu^a$ and $\omega^a$, whereas 
 the equations (\ref{a1})--(\ref{a3}) imply that the dependence of 
 $(\Omega^{\mu a},\bar c^a)$, $(\ga^{\mu\nu a},\bar\xi^a_\mu)$ and $(\ro^{\mu a},\bar\phi^a)$
 is given by the following combinations
 \pe \tilde\Omega^{\mu a} &=& \Omega^{\mu a}-\sqrt gg^{\mu\al}n_\al\bar c^a \ptr, \nm \\
     \tilde\ga^{\mu\nu a} &=& \ga^{\mu\nu a}-\sqrt gg^{\mu\al}g^{\nu\be}(n_\al\bar\xi^a_\be-
     n_\be\bar\xi_\al^a) \ptr, \nm \\
     \tilde\ro^{\mu a} &=& \ro^{\mu a}+\sqrt gg^{\mu\al}n_\al\bar\phi^a\ptr. \se
 The equations (\ref{con1})--(\ref{con2}), as in reference \cite{Joe}, can be 
 unified into a single operator $\da$:
 \en \da={\cal S}_{\Sigma}+{\cal W}^S_{(\eta)}+{\cal W}^D_{(\ee)}+\int d^4x\left\{
     \left[\ee,\eta\right]^\mu{\da\over\da\eta^\mu}+\left({1\over 2}\left[\ee,\ee\right]^\mu
     -\eta^\mu\right){\da\over\da\ee^\mu}\right\}\dv
 producing a cohomology problem
 \en \da\Delta=0 \ptr. \label{cohproblem} \dv
 It can be easily verified that the operator $\delta$ is nilpotent
 \en \da^2=0 \ptr. \dv
 Therefore, any expression  of the form
 $\da\hat\Delta$ is automatically a solution of (\ref{cohproblem}). A solution of 
 this type is called a trivial solution. 
 Hence, the most general solution of (\ref{cohproblem}) reads 
 \en \Delta=\Delta_c+\da\hat\Delta \ptr. \dv
 Here, the nontrivial solution $\Delta_c$ is $\da$-closed $(\da\Delta_c=0)$, but not trivial 
 $(\Delta_c\ne\da\hat\Delta)$. 
 \\[2mm]
 We begin with 
 the determination of the nontrivial solution of (\ref{cohproblem}). For this purpose
 we introduce a filtering operator ${\cal N}$:
 \en {\cal N}=\int d^4x\,\sum_{\varphi}\varphi{\da\over\da\varphi} \ptr, \dv
 where $\varphi$ stands for all fields, including $n_\mu,\chi_\mu,\ee^\mu$ and $\eta^\mu$.  
 To all fields we assign the homogeneity degree 1. 
 The filtering operator induces a decomposition
 of $\da$ according to
 \en \da=\da_0+\da_1 \ptr. \label{decomp}\dv
 The operator $\da_0$ does not increase the homogeneity degree while acting on a field
 polynomial. On the other hand, the operator $\da_1$ increases the homogeneity degree by
 one unit. 
 Furthermore, the nilpotency of $\da$ leads to 
 \en \da_0^2=0\pr,\pp \{\da_0,\da_1\}=0\pr,\pp \da_1^2=0 \ptr. \label{relat} \dv
 Hence, we obtain from (\ref{relat}) the following relation
 \en \da_0\Delta=0\ptr, \dv
 which yields a further cohomology problem.
 The usefulness of the decomposition  
 (\ref{decomp}) relies on a very general theorem \cite{Piguet} stating that the cohomology of the
 complete operator $\da$ is isomorphic to a subspace of the cohomology of the operator $\da_0$.
The cohomology of $\da_0$ is easier to solve than the cohomology of $\da$.
 The operator $\da_0$ acts on the fields as follows:
 \en
 \begin{tabular}{ll}
  $\da_0 A_\mu^a=-\dms c^a \ptr,$ & $\da_0 B_{\mu\nu}^a=-\dms\xi_\nu^a+\de_\nu\xi_\mu^a$\ptr,\\
  $\da_0 c^a=0 \ptr,  $& $\da_0\xi_\mu^a=\dms\phi^a \ptr, $ \\
  $\da_0\phi^a=0 \ptr, $& $\da_0\tilde\ro^{\mu a}=\de_\nu\tilde\ga^{\mu\nu a}\ptr,$ \\
  $\da_0\tilde\Omega^{\mu a}={1\over 2}\ee^{\mu\nu\ro\si}\de_\nu B_{\ro\si}^a$\ptr, & 
  $\da_0\tilde\ga^{\mu\nu a}=\ee^{\mu\nu\ro\si}\de_\ro A_\si^a$ \ptr, \\
  $\da_0 L^a=-\dms\tilde\Omega^{\mu a}$\ptr,  & $\da_0 D^a=-\dms\tilde\ro^{\mu a} \ptr,  $  \\
  $\da_0 g_{\mu\nu}=\hat g_{\mu\nu}$\ptr, & $\da_0\hat g_{\mu\nu}=0 \ptr, $ \\
  $\da_0 n_\mu=\chi_\mu$\ptr, & $\da_0\chi_\mu=0\ptr, $  \\
  $\da_0\ee^{\mu}=-\eta^\mu \ptr,$ & $\da_0\eta^\mu=0 \ptr, $  \ptr.
   \end{tabular}   \label{delta0}
 \dv
 We notice that the quantities $g_{\mu\nu},\hat g_{\mu\nu},n_\mu,\chi_\mu,\ee^\mu$ and $\eta^\mu$
 transform under $\da_0$
 as doublets, being therefore out of the cohomology \cite{Piguet,Brandt}. 
 The nontrivial solution
 $\Delta_c$ can now be written as integrated local field polynomial of form degree four and
 ghost number zero:
 \en \Delta_c=\int \omega^0_4 \ptr, \dv
 where $\omega^p_q$ is a field polynomial of form degree $q$ and ghost number $p$. 
 Using the Stoke's theorem, the algebraic Poincar\'e lemma \cite{Brandt} and the relation 
 $\{\da_0,d\}=0$, where $d$ represents the nilpotent exterior derivative ($d^2=0$),
 we obtain the following tower of descent equations:
 \pe \da_0\omega^0_4+d\omega_3^1 &=& 0 \ptr, \nm \\
     \da_0\omega_3^1+d\omega^2_2 &=& 0 \ptr, \nm \\
     \da_0\omega_2^2+d\omega^3_1 &=& 0 \ptr, \nm \\
     \da_0\omega_1^3+d\omega^4_0 &=& 0 \ptr, \nm \\
     \da_0\omega^4_0 &=&0 \ptr. \label{tower} \se
 The tower of descent equations (\ref{tower}) has been solved in \cite{Joe}, where it
 was shown that $\omega_0^4$ takes the following form:
 \en \omega_0^4=u\phi^a\phi^a+vf^{abc}c^ac^b\phi^c \ptr, \dv
 with $u$ and $v$ being some constant coefficients. In \cite{Joe}, the authors showed by using
the equation (\ref{ghost234}) that both $u$ and $v$ vanish.
 \\[2mm]
 Next, we move to the computation of the trivial counterterms which are constrained by the
 dimension, the ghost number and the weight requirements. The most general trivial solution
 can be constructed as follows
 \pe \hat\Delta &=& \int d^4x\left(\al_1\tilde\Omega^{\mu a}\am^a+\right.
     \al_2\tilde\ga^{\mu\nu a}B_{\mu\nu}^a+\al_3L^ac^a+\al_4\tilde\ro^{\mu a}\xi_\mu^a+  
     \al_5D^a\phi^a+ \nm \\
 &+& \al_6f^{abc}\tilde\ga^{\mu\nu a}\am^bA_\nu^c+\al_7f^{abc}\tilde\ro^{\mu a}
     \am^bc^c+\al_8\tilde\ga^{\mu\nu a}\dms A_\nu^a+\al_9\tilde\ro^{\mu a}A_\nu^ag^{\ro\nu}
     \hat g_{\ro\mu}+ \nm \\
 &+& \al_{10}f^{abc}D^ac^bc^c+\al_{11}f^{abc}\ee_{\mu\nu\ro\si}\tilde\ga^{\mu\nu a}
     \tilde\ga^{\ro\si b}c^c+\al_{12}D^ac^a\hat g_{\mu\nu}g^{\mu\nu}+\al_{13}\tilde\ro^{\mu a}
     \dms c^a+ \nm \\
 &+& \al_{14}{1\over\sqrt g}\tilde\ga^{\mu\nu a}\tilde\ga^{\ro\si a}\hat g_{\mu\ro}g_{\nu\si}+
     \al_{15}\sqrt g\ee_{\mu\nu\ro\si}\tilde\ga^{\mu\nu a}g^{\ro\al}g^{\si\be}\de_\al A_\be^a+
     \nm \\
 &+& \al_{16}{1\over\sqrt g}f^{abc}g_{\mu\ro}g_{\nu\si}\tilde\ga^{\mu\nu a}\tilde\ga^{\ro\si b}
     c^c+\al_{17}\tilde\ro^{\mu a}\am^ag^{\ro\si}\hat g_{\ro\si}+  \nm \\
 &+& \al_{18}\sqrt g\ee_{\mu\nu\ro\si}f^{abc}\tilde\ga^{\mu\nu a}g^{\ro\al}g^{\si\be}A_\al^b
     A_\be^c+\al_{19}\sqrt g\ee_{\mu\nu\ro\si}g^{\ro\al}g^{\si\be}\tilde\ga^{\mu\nu a}
     B_{\al\be}^a+ \nm \\
 &+& \al_{20} n_\mu\tilde\Omega^{\mu a}g^{\ro\al}n_\al A_\ro^a+
     \al_{21} g^{\mu\al}n_\al n_\ro B_{\mu\be}^a\tilde\ga^{\be\ro a}+
     \al_{22} n_\mu\tilde\ro^{\mu a}g^{\nu\be}n_\be\xi_\nu^a+ \nm \\
 &+& \al_{23} n_\mu g^{\ro\al}n_\al f^{abc}\tilde\ga^{\mu\nu a}A_\nu^bA_\ro^c+
     \al_{24} n_\mu g^{\ro\al}n_\al f^{abc}\tilde\ro^{\mu a}A_\ro^bc^c+
     \al_{25} n_\mu\tilde\ga^{\mu\nu a}\de_\nu(g^{\ro\al}n_\al A_\ro^a)+ \nm \\
 &+& \al_{26} n_\mu\tilde\ro^{\mu a}g^{\nu\al}n_\al A_\nu^ag^{\ro\si}\hat g_{\ro\si}+
     \al_{27} n_\mu\tilde\ro^{\mu a}g^{\ro\nu}A_\nu^a\hat g_{\ro\tau}g^{\tau\al}n_\al+
     \al_{28} \tilde\ro^{\mu a}n_\nu g^{\nu\al}A_\al^a\hat g_{\mu\tau}g^{\tau\be}n_\be+\nm \\
 &+& \al_{29} D^ac^ag^{\mu\nu}n_\mu\hat g_{\nu\tau}g^{\tau\al}n_\al+
     \al_{30} n_\mu\tilde\ro^{\mu a}g^{\al\nu}\de_\nu(c^an_\al)+
     \al_{31}{1\over\sqrt g}\tilde\ga^{\mu\nu a}\tilde\ga^{\ro\si a}n_\mu\hat g_{\nu\ro}n_\si+
     \nm \\
 &+& \al_{32} \sqrt g\ee_{\mu\nu\ro\si}\tilde\ga^{\mu\nu a}g^{\ro\al}g^{\si\tau}g^{\be\la}
          n_\tau n_\la\de_\al A_\be^a+
     \al_{33} \sqrt g\ee_{\mu\nu\ro\si}\tilde\ga^{\mu\nu a}g^{\ro\tau}g^{\al\la}g^{\si\be}
          n_\tau n_\la\de_\al A_\be^a+ \nm \\
 &+& \al_{34} \sqrt g\ee_{\mu\nu\ro\si}g^{\mu\al}n_\al\tilde\ga^{\nu\tau a}n_\tau 
          g^{\ro\be}g^{\si\la}\de_\be A_\la^a+ 
     \al_{35} {1\over\sqrt g}f^{abc}\tilde\ga^{\mu\nu a}\tilde\ga^{\ro\si b}c^c
          n_\mu n_\ro g_{\nu\si}+ \nm \\
 &+& \al_{36} \tilde\ro^{\mu a}n_\mu g^{\nu\al}n_\al A_\nu^ag^{\ro\si}\hat g_{\ro\si}+ 
     \al_{37} \tilde\ro^{\mu a}n_\mu g^{\nu\al}n_\al A_\nu^ag^{\ro\be}n_\be
          g^{\si\la}n_\la\hat g_{\ro\si}+ \nm \\      
 &+& \al_{38} \tilde\ro^{\mu a}A_\mu^ag^{\ro\al}n_\al g^{\si\be}n_\be\hat g_{\ro\si}+ 
     \al_{39} \sqrt g\ee_{\mu\nu\ro\si}f^{abc}\tilde\ga^{\mu\nu a}g^{\ro\al}A_\al^b
          g^{\si\be}n_\be g^{\la\tau}n_\tau A_\la^c+ \nm \\
 &+& \al_{40} \sqrt g\ee_{\mu\nu\ro\si}f^{abc}g^{\mu\al}n_\al\tilde\ga^{\nu\tau a}n_\tau
          g^{\ro\be}g^{\si\la}A_\be^b A_\la^c+
     \al_{41} \sqrt g\ee_{\mu\nu\ro\si}\tilde\ga^{\mu\nu a}g^{\ro\al}n_\al g^{\si\be}
          g^{\la\tau}n_\tau B_{\la\be}^a+ \nm \\                                    
 &+& \al_{42} \sqrt g\ee_{\mu\nu\ro\si}g^{\mu\la}n_\la\tilde\ga^{\nu\tau a}n_\tau 
          g^{\ro\al}g^{\si\be}B_{\al\be}^a+
     \al_{43} {1\over\sqrt g}g_{\mu\ro}g_{\nu\si}\tilde\ga^{\mu\nu a}\tilde\ga^{\ro\si a}
     g^{\al\be}n_\al\chi_\be+
     \nm \\
 &+& \al_{44} {1\over\sqrt g}g_{\mu\be}\tilde\ga^{\al\mu a}n_\al\tilde\ga^{\be\nu a}n_\nu
          g^{\ro\la}n_\la\chi_\ro+
     \al_{45} g^{\mu\al}n_\al\chi_\mu D^ac^a+
     \al_{46} g^{\mu\al}n_\al\chi_\mu\tilde\ro^{\ro a}A_\ro^a + \nm \\
 &+& \al_{47} \ee_{\mu\nu\ro\si}g^{\mu\al}n_\al g^{\nu\be}\chi_\be g_{\tau\la}
          \tilde\ga^{\ro\tau a}\tilde\ga^{\la\si a}+
     \al_{48} \ee_{\mu\nu\ro\si}g^{\mu\al}n_\al g^{\nu\be}\chi_\be 
          \tilde\ga^{\ro\tau a}n_\tau\tilde\ga^{\si\la a}n_\la + \nm \\
 &+& \al_{49} \sqrt{g}\ee_{\mu\nu\ro\si}g^{\mu\al}n_\al g^{\nu\be}\chi_\be\tilde\ro^{\ro a}g^{\si\la}
          A_\la^a +
     \left.\al_{50} \tilde\ro^{\mu a}n_\mu g^{\nu\al}\chi_\al A_\nu^a\right)\ptr. \label{alfas} \se                                        
 The trivial counterterm\footnote{In (\ref{alfas}) the quantities $\al_i, i=1,\ldots,50$ are constant
 coefficients to be determined.} may depend on the quantities $\eta^\mu$ and $\ee^\mu$
 which do not appear in the total action (\ref{Sigma}). For this reason we demand the
 expression $\da\hat\Delta$ to be independent of the parameters $\eta^\mu$ and $\ee^\mu$. 
 In fact, after a tedious computation $\hat\Delta$ reduces to an expression 
which is forbidden by (\ref{ghost234}). Thus, all of the coefficients $\al_i, i=1,\ldots,50$
vanish. \\
 Therefore, we have shown that the total action $\Sigma$ does not admit any deformations at
the quantum level. 
 \\[2mm]
 The last step in our analysis is devoted to the discussion of the existence of possible breaking of the
 symmetries at the quantum level. 
By using the same arguments as in \cite{Joe} and under the assumption
 that the quantum action principle is also valid in the case of non-covariant gauges
 \cite{Buch}, one can easily show that
the symmetries of the model do not admit any anomalies and therefore,
 are valid at the quantum level. This completes the proof of finiteness of the four dimensional
antisymmetric tensor field model to all orders of perturbation theory, quantized on a topologically
trivial and asymptotically flat manifold.


\begin{thebibliography}{99}
 \bibitem{Joe} U. Feichtinger, O. Moritsch, J. Rant, M. Schweda and H. Zerrouki,
   {\sl Int. J. Mod. Phys.} {\bf A13} (1998) 4513-4538;

\bibitem{Hassan} H. Zerrouki, {\sl Il Nuovo Cimento} {\bf 112A} (1999) 395;
   
\bibitem{Piguet} O. Piguet and S. P. Sorella, {\sl Algebraic Renormalization, Perturbative
   Renormalization, Symmetries and Anomalies}, Lecture Notes in Physics, New Series m: 28,
   Springer Verlag 1995;

\bibitem{Buch} A. Boresch, S. Emery, O. Moritsch, M. Schweda, T. Sommer and H. Zerrouki,
   {\sl Applications of noncovariant gauges in the algebraic renormalization procedure},
   World Scientific 1998;
 
\bibitem{Becchi} C. M. Becchi, A. Rouet, R. Stora, {\sl Comm. Math. Phys.} {\bf 42} (1975) 
   127; \\ {\sl Ann. Phys. (N.Y.)} {\bf 98} (1976) 287;
 
\bibitem{ndep2} O. Piguet and K. Sibold, {\sl Nucl. Phys.} {\bf B253} (1985) 517;
          
   
 \bibitem{Brandt} F. Brandt, N. Dragon, M. Kreuzer, {\sl Phys. Lett.} {\bf B231} (1989) 263;
   {\sl Nucl. Phys.} {\bf B332} (1990) 224; {\sl Nucl. Phys.} {\bf B340} (1990) 187;
 
 
 
   
 \end{thebibliography}
\end{document}